\documentclass[twocolumn,showpacs,aps,amsmath,amssymb,prl,superscriptaddress]{revtex4-2}  
\usepackage[latin9]{inputenc}
\usepackage{textcomp}
\usepackage{amsmath}
\usepackage{graphicx}
\PassOptionsToPackage{version=3}{mhchem}
\usepackage{mhchem}

\makeatletter

\usepackage{subfigure}\usepackage{epstopdf}\usepackage{dcolumn}
\usepackage{bm}
\usepackage{multirow}
\usepackage{longtable}\usepackage{rotating}\usepackage{threeparttable}
\usepackage{float}%Position vector
\def\br{{\bf r}}
\def\bx{{\bf x}}
\makeatother

\begin{document}
\title{Unity of Kohn-Sham Density Functional Theory and Reduced Density Matrix Functional Theory}
\author{Neil Qiang Su}
\email{nqsu@nankai.edu.cn}
\affiliation{Department of Chemistry, Key Laboratory of Advanced Energy Materials Chemistry (Ministry of Education) and Renewable Energy Conversion and Storage Center (RECAST), Nankai University, Tianjin 300071, China}

\begin{abstract}
This work presents a theory to unify the two independent theoretical frameworks of Kohn-Sham (KS) density functional theory (DFT) and reduced density matrix functional theory (RDMFT). The generalization of the KS orbitals to hypercomplex number systems leads to the hypercomplex KS (HCKS) theory, which extends the search space for the density in KS-DFT to a space that is equivalent to natural spin orbitals with fractional occupations in RDMFT. Thereby, HCKS is able to capture the multi-reference nature of strong correlation by dynamically varying fractional occupations. Moreover, the potential of HCKS to overcome the fundamental limitations of KS is verified on systems with strong correlation, including atoms of transition metals. As a promising alternative to the realization of DFT, HCKS opens up new possibilities for the development and application of DFT in the future.
\end{abstract}
\maketitle

$Introduction.$---Built upon the Hohenberg-Kohn theorem \cite{HK1964,Levy1979pnas}, Kohn-Sham (KS) density functional theory (DFT) \cite{KS1965,PY1989,Dreizler2012} is a formally exact theoretical framework toward the many-electron problem. Due to the favorable balance between accuracy and efficiency, KS-DFT has won enormous popularity that can manifest itself in the countless applications across physics, materials science, chemistry, and biology \cite{Peverati2014trc,Jones2015,Sun2016nc}. Nonetheless, the great success of KS-DFT, along with commonly used density functional approximations (DFAs), is clouded by the improper treatment of strong correlation \cite{Cohen2008science}.
Strong correlation represents the intractable electronic interaction stemming from the multi-reference nature of systems, which has long posed a major challenge to KS-DFT \cite{Burke2012jcp,Cohen2012cr,Becke2014jcp,Su2017arpc}.

It is generally recognized that intrinsic errors associated with commonly used DFAs \cite{Ruzsinszky2006jcp,Ruzsinszky2007jcp,Vydrov2007jcp,Paula2008prl,Paula2009prl,Cohen2008science,Zheng2011,Chai2013prl}  and the slow progress in systematically eliminating these errors to better handle strong correlation have severely limited the applicability of KS-DFT \cite{Paula2009prl,Cohen2008science,Mardirossian2017,Su2018pnas}.
The enlightening work by Lee, Bertels, Small, and Head-Gordon \cite{Lee2019prl} shows that commonly used DFAs can better treat the strong correlation in singlet biradicals by using complex spin-restricted orbitals in KS-DFT. This thus raises a deep question: In addition to the errors inherent in existing DFAs, is there still any limitation in the understanding and application of the KS-DFT framework? Apparently, in-depth insights into this question is important for further development and application of DFT.

Beside KS-DFT, reduced density matrix functional theory (RDMFT) \cite{Gilbert1975prb,Muller1984rpa,GU1998prl,Pernal2005prl,Sharma2008prb,Piris2010jcp,Sharma2013prl,Schade2017,Schilling2019prl} provides an alternative approach to the many-electron problem. The Gilbert's theorem \cite{Gilbert1975prb} guarantees that the one-electron reduced density matrix (1-RDM) instead of the density can be used as the fundamental variable for the energy functional, which thus makes RDMFT an exact theoretical framework independent of KS-DFT. RDMFT has proved its great potential to overcome the fundamental limitations of KS-DFT through the successful application in predicting dissociation energy curves \cite{Gritsenko2005jcp,Rohr2008jcp,Lathiotakis2009pra}, and fundamental gaps for finite systems and extended solids \cite{Lathiotakis2010zpc} as well as for Mott insulators \cite{Sharma2008prb}. Therefore, establishing a connection with the theoretical framework of RDMFT will further improve our understanding of KS-DFT and the problems of existing DFAs.

This letter seeks to come up with a theory to unify the two theoretical frameworks of KS-DFT and RDMFT. This is achieved by generalizing the conventional KS determinant to hypercomplex number systems. The resulting hypercomplex KS (HCKS) theory extends the search space for the density in KS-DFT to a space that is equivalent to natural spin orbitals with fractional occupations in RDMFT. The potential of HCKS in capturing the physical essence of strong correlation is demonstrated on atoms of multi-reference character, including transition metals.

$Theory.$---The total energy $E_{\rm{tot}}[\rho_\sigma]$ in KS-DFT includes the kinetic energy $T_{\rm{s}}[\rho_\sigma]$, the external energy $E_{\rm{ext}}[\rho_\sigma]$, the Coulomb energy $E_{\rm{H}}[\rho_\sigma]$, and the exchange-correlation (XC) energy $E_{\rm{XC}}[\rho_\sigma]$. They are all functionals of the density that can be formed by the occupied KS orbitals in the KS determinant ($N_\sigma$ being the $\sigma$-spin electron number),
 \begin{equation}
\label{eq:rho_ks}
\rho_\sigma(\br)= \sum_{k=1}^{N_\sigma} |\varphi_k^\sigma(\br)|^2,
\end{equation}
while $T_{\rm{s}}[\rho_\sigma]$ can be explicitly formulated as $-\frac{1}{2}\sum_{\sigma}^{\alpha,\beta}\sum_{k=1}^{N_\sigma} \langle \varphi_{k}^{\sigma} | \nabla^2 | \varphi_{k}^{\sigma} \rangle$ \cite{KS1965}.
Therefore, the minimization of $E_{\rm{tot}}[\rho_\sigma]$ with respect to $\rho_\sigma$ is equivalent to the minimization with respect to $\{\varphi_p^\sigma\}$, subject to the orthonormalization condition
\begin{equation}
\label{eq:ortho_ks}
\langle \varphi_p^\sigma | \varphi_q^\sigma \rangle=\delta_{pq}.
\end{equation}

In RDMFT, the four terms in the total energy are uniquely determined by 1-RDM ($\gamma_\sigma$), which are $T[\gamma_\sigma]$, $E_{\rm{ext}}[\gamma_\sigma]$, $E_{\rm{H}}[\gamma_\sigma]$ and $E_{\rm{XC}}[\gamma_\sigma]$ respectively. In terms of the natural spin orbitals $\{\psi_p^\sigma\}$ and the natural occupation numbers $\{n_p^\sigma\}$, $\gamma_\sigma$ in the spectral representation reads $\sum_{p=1}^{K}|\psi_p^\sigma \rangle n_p^\sigma \langle\psi_p^\sigma|$ ($K$ being the dimension of the basis set), and the diagonal element $\langle \br | \gamma_\sigma | \br \rangle$ is the density,
\begin{equation}
\label{eq:rho_rdmft}
\rho_\sigma(\br)=\sum_{p=1}^{K} n_p^\sigma |\psi_p^\sigma(\br)|^2.
\end{equation}
$T[\gamma_\sigma]$ has the form, $-\frac{1}{2}\sum_{\sigma}^{\alpha,\beta}\sum_{p=1}^{K}n_p^\sigma \langle \psi_{p}^{\sigma} | \nabla^2 | \psi_{p}^{\sigma} \rangle$.
The ground-state energy can be obtained by the minimization of $E_{\rm{tot}}[\gamma_\sigma]$ with respect to $\gamma_\sigma$, or equivalently, with respect to both $\{\psi_p^\sigma\}$ and $\{n_p^\sigma\}$, subject to the orthonormalization condition, $\langle \psi_p^\sigma | \psi_q^\sigma \rangle=\delta_{pq}$, and the Pauli exclusion principle and the $N$-representability constraint \cite{Coleman1963rmp},
\begin{equation}
\label{eq:occ_rdmft}
0\leq n_p^\sigma \leq 1, \sum_{p=1}^K n_p^\sigma=N_\sigma.
\end{equation}
Unlike KS-DFT, RDMFT allows dynamically varying fractional occupations to capture the multi-reference nature of strong correlation.

The connection between KS-DFT and RDMFT is established by introducing hypercomplex. The concept of hypercomplex \cite{HC1989}, and the theory of Clifford algebra that generalizes real numbers, complex numbers to quanternions, octonions, and other hypercomplex numbers have important applications in a variety of fields including theoretical physics \cite{CA2019}. Here, the HCKS orbitals are formulated as
\begin{equation}
\label{eq:orb_hc}
\varphi_p^\sigma(\br)=\phi_{p}^{\sigma,0}(\br)+\sum_{\mu=1}^{n}\phi_{p}^{\sigma,\mu}(\br) e_\mu,
\end{equation}
where $\{\phi_{p}^{\sigma,\mu}\}$ are a set of real functions, and $\{e_1, e_2, \cdots, e_n\}$ are a basis of dimension $n$ in a Clifford algebra, such that \cite{CA2019}
\begin{equation}
\label{eq:CA}
e_\mu^2=-1; e_\mu e_\nu=-e_\nu e_\mu.
\end{equation}
The conjugate hypercomplex of the HCKS orbitals are $\bar{\varphi}_p^\sigma(\br)=\phi_{p}^{\sigma,0}(\br)-\sum_{\mu=1}^{n}\phi_{p}^{\sigma,\mu}(\br) e_\mu$. Therefore, Eq \ref{eq:orb_hc} provides a general set of high dimensional orbitals for the HCKS determinant, while complex KS orbitals are a special case for $n=1$.

Without loss of generality, $\{\phi_{p}^{\sigma,\mu}\}$ can be expanded on a set of orthonormal functions $\{\chi_p\}$ and read
\begin{equation}
\label{eq:chi2phi}
\phi_{p}^{\sigma,\mu}(\br)=\sum_{q=1}^K \chi_q(\br)V_{pq}^{\sigma,\mu}.
\end{equation}
Here ${\rm{V}}^{\sigma,\mu}$ is a $K\times K$ matrix associated with the $\mu$-th component of the HCKS orbitals. The orthonormalization condition of Eq \ref{eq:ortho_ks} becomes \cite{footnote}
\begin{equation}
\label{eq:cond1_hc}
\sum_{\mu=0}^{n} {\rm{V}}^{\sigma,\mu}{\rm{V}}^{\sigma,\mu T}=\sum_{\mu=0}^{n} {\rm{V}}^{\sigma,\mu T}{\rm{V}}^{\sigma,\mu}={\rm{I}}_K,
\end{equation}
and
\begin{align}
\label{eq:cond2_hc}
\begin{cases}
{\rm{V}}^{\sigma,\mu}{\rm{V}}^{\sigma,\nu T}={\rm{V}}^{\sigma,\nu}{\rm{V}}^{\sigma,\mu T}; \\
{\rm{V}}^{\sigma,\mu T}{\rm{V}}^{\sigma,\nu}={\rm{V}}^{\sigma,\nu T}{\rm{V}}^{\sigma,\mu},
\end{cases}
\end{align}
where the superscript $T$ denotes the transpose, and ${\rm{I}}_K$ is the $K\times K$ identity matrix. The density corresponding to the HCKS determinant reads \cite{footnote},
\begin{equation}
\label{eq:Drho}
\rho_\sigma(\br)=\sum_{p,q = 1}^K \chi_p(\br) D_{pq}^{\sigma} \chi_q(\br),
\end{equation}
and ${\rm{D}}^\sigma$ is defined by
\begin{equation}
\label{eq:D_hc}
{\rm{D}}^\sigma=\sum_{\mu=0}^{n} {\rm{V}}^{\sigma,\mu T}{\rm{I}}_K^{N_\sigma}{\rm{V}}^{\sigma,\mu},
\end{equation}
where ${\rm{I}}_K^{N_\sigma}$ is a $K \times K$ diagonal matrix, with the first $N_\sigma$ diagonal elements being 1 and the rest being 0. ${\rm{D}}^\sigma$ is symmetric and can be diagonalized by an orthogonal matrix ${\rm{U}}^\sigma$, 
\begin{equation}
\label{eq:ULU}
{\rm{D}}^\sigma={\rm{U}}^\sigma {\rm{\Lambda}}^\sigma {\rm{U}}^{\sigma T},
\end{equation}
where ${\rm{\Lambda}}^\sigma$ is a diagonal matrix, diag$(\lambda_1^\sigma, \lambda_2^\sigma, \cdots, \lambda_K^\sigma)$, and $\{\lambda_p^\sigma\}$ are the eigenvalues of ${\rm{D}}^\sigma$, which satisfy \cite{footnote}
\begin{equation}
\label{eq:occ_cond}
0 \leq \lambda_p^\sigma \leq 1; \sum_{p=1}^K \lambda_p^\sigma=N_\sigma.
\end{equation}
By inserting Eq \ref{eq:ULU}, the density of Eq \ref{eq:Drho} can be written as
\begin{equation}
\label{eq:Vrho}
\rho_{\sigma}(\br) = \sum_{p=1}^{K}{\lambda}_{p}^{\sigma} |\chi_p^{\sigma}(\br)|^2,
\end{equation}
where $\chi_p^{\sigma}(\br)=\sum_{q=1}^K \chi_q(\br)U_{qp}^{\sigma}$ (thereby $\{\chi_k^\sigma\}$ being orthonormal). Similarly, the kinetic energy reads
\begin{equation}
\label{eq:VT}
T_{\rm{s}}[\rho_\sigma]=-\frac{1}{2}\sum_{\sigma}^{\alpha,\beta}\sum_{p=1}^{K} \lambda_{p}^{\sigma} \langle \chi_p^\sigma | \nabla^2 | \chi_p^{\sigma} \rangle.
\end{equation}

To further verify that $\{\lambda_p^\sigma\}$ in Eqs \ref{eq:Vrho} and \ref{eq:VT}, associated with any given orthonormal functions $\{\chi_k^\sigma\}$, can take any values subject to Eq \ref{eq:occ_cond} when $K\leq n+1$, here construct a special set of HCKS orbitals. Given a set of orthonormal $\{\chi_k^\sigma\}$ , $\{\phi_p^{\sigma,\mu}\}$ of Eq \ref{eq:chi2phi} are constructed with the following form
\begin{equation}
\label{eq:phi2}
\phi_p^{\sigma,\mu}= \chi_{\mu+1}^\sigma(\br) V_{p \mu+1}^{\sigma,\mu},
\end{equation}
i.e. the components corresponding to the same $e_\mu$ in all the HCKS orbitals are formed by the same function $\chi_{\mu+1}^\sigma$, thereby
\begin{align}
\label{eq:V2}
V_{pq}^{\sigma,\mu}=\begin{cases}
W_{pq}^\sigma, & q=\mu+1\\
0, & q\neq \mu+1,
\end{cases}
\end{align}
where ${\rm{W}}^\sigma$ is a $K \times K$ matrix. The orthonormalization condition in terms of ${\rm{W}}^\sigma$ reads
\begin{equation}
\label{eq:cond11_hc}
{\rm{W}}^{\sigma}{\rm{W}}^{\sigma T}={\rm{I}}_K,
\end{equation}
which thus requires ${\rm{W}}^\sigma$ to be an orthogonal matrix. Similar derivations lead to Eqs \ref{eq:Vrho} and \ref{eq:VT} for $\rho_\sigma(\br)$ and $T[\rho_\sigma]$ respectively, with $\lambda_p^\sigma = \sum_{k=1}^{N_\sigma}(W_{kp}^{\sigma})^2$. As ${\rm{W}}^\sigma$ can be any orthogonal matrix, $\{\lambda_p^\sigma\}$ can take any values subject to Eq \ref{eq:occ_cond}. 

Therefore, the extension of KS-DFT to hypercomplex orbitals leads to a density that has the same search space as the density of Eq \ref{eq:rho_rdmft} constructed by real orbitals in RDMFT, with the same form of kinetic energies as well. The resulting HCKS theory thus unifies KS-DFT and RDMFT, which extends the search space of KS-DFT to natural spin orbitals with fractional occupations. When $n=0$, the orbitals of Eq \ref{eq:orb_hc} are real and HCKS reduces the conventional KS method; when $n=1$, the orbitals are complex, which are used in the complex, spin-restricted KS (CRKS) method \cite{Lee2019prl}.

$Results.$---The numerical performance of HCKS and its potential in describing strong correlation are evaluated with the finite basis simulation. Atoms of multi-reference character, including transition metals, are calculated. While two XC functionals, PBE \cite{PBE96} and BLYP \cite{B88,LYP}, are examined here, similar results can be obtained with other commonly used functionals. All calculations were performed using a local modified version of the NWChem package \cite{NWChem}.

\begin{figure}[htbp]
 \centering
 \includegraphics[width=1\linewidth]{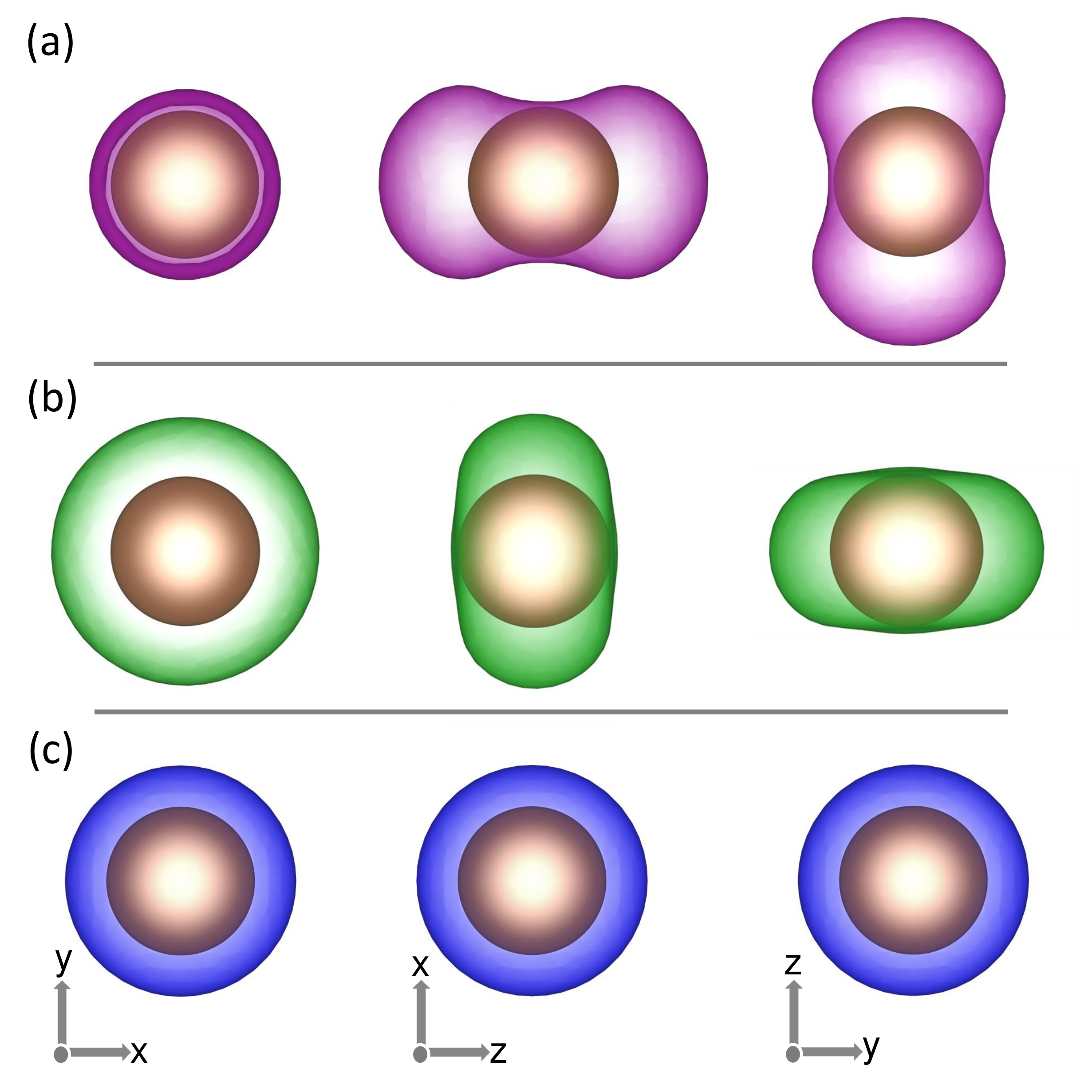}
 \caption{Density on real-space grids for the lowest singlet state of C atom.
Views along the z, y and x axes are provided respectively. 
(a) RKS density (purple), (b) CRKS density (green), and (c) HCKS (or equivalently, RDMFT) density (blue) are plotted at the isosurface value of 0.2 au. The XC functional PBE is applied for all the calculation. The basis set used is aug-cc-pVQZ \cite{Dunning1989jcp,Kendall1992jcp}.}
 \label{fig:Density} 
\end{figure}

The density for the lowest singlet state of C atom is first examined. The singlet-state C atom is of biradical nature, with two electrons in the 2p sub-shell. Normally, the spin-restricted KS (RKS) would have one doubly occupied p orbital and destroy the degeneracy of the p orbitals. Fig \ref{fig:Density}a shows the density of RKS when p$_{\rm{z}}$ is occupied, which maintains the spin symmetry but loses the space symmetry. Different from RKS, CRKS leads to two half filled p orbitals for the singlet C atom \cite{Lee2019prl,footnote}. Fig \ref{fig:Density}b shows that the CRKS density with both p$_{\rm{x}}$ and p$_{\rm{y}}$ half filled maintains the space symmetry along the z axis, but it cannot guarantee the apace symmetry along both x and y axes. In contrast, HCKS, or equivalently, RDMFT, automatically converges to the spin-restricted solution, with each p orbital 1/3 filled. Therefore, the density by HCKS can maintain both spin and space symmetry; see Fig \ref{fig:Density}c. Similar results can be obtained for other systems of multi-reference character such as singlet states of O, S, and Si atoms, and even the d and f block transition metals.

\begin{figure}[htbp]
 \centering
 \includegraphics[width=1\linewidth]{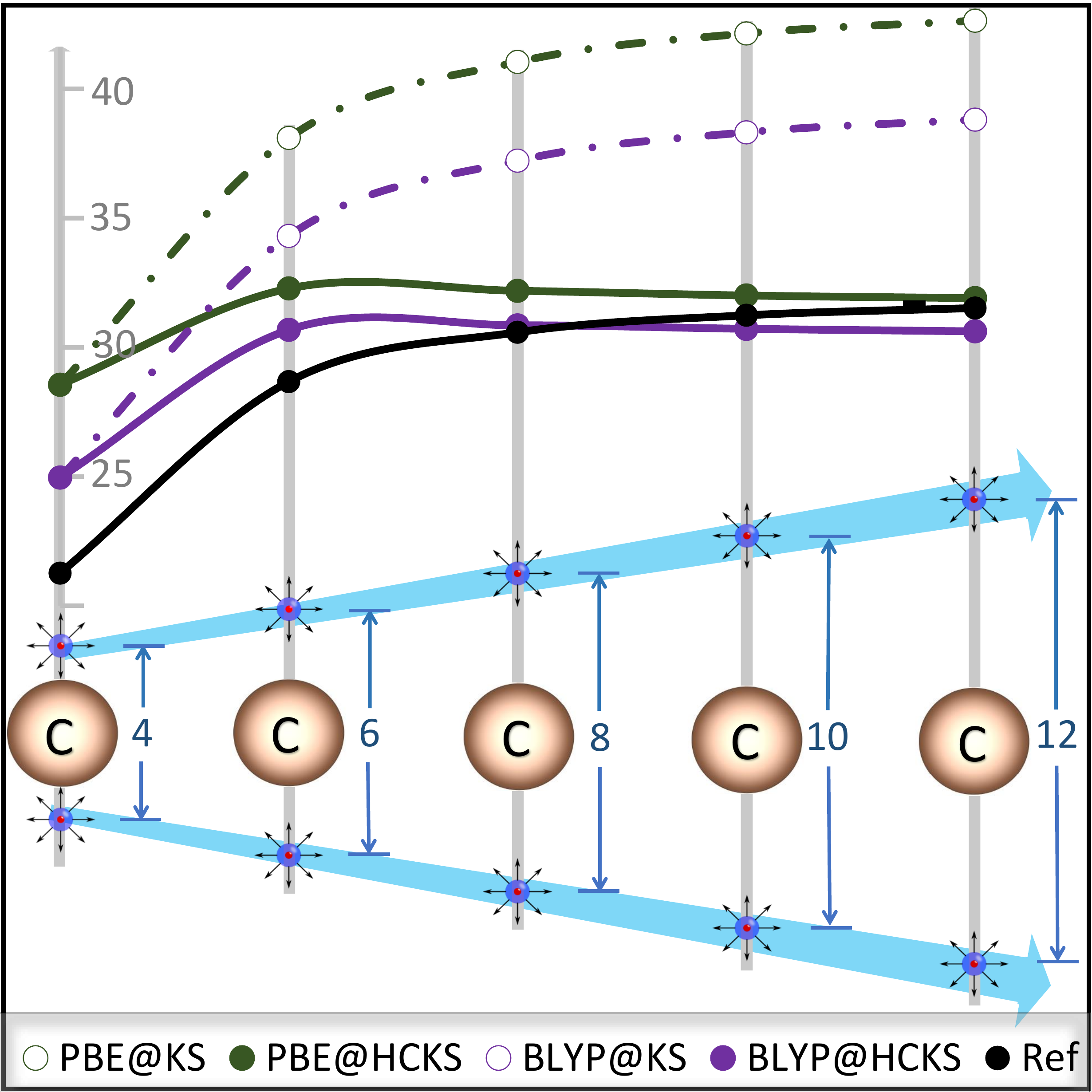}
 \caption{Triplet-singlet energy gaps of C atom under varying external charged environments. The energies are calculated with two point charges of 0.3 au each placed equidistantly on two sides of the C atom, with the distance between the two point charges ranging from 4 to 12 \AA. KS and HCKS (or equivalently, RDMFT) in use of the same XC functionals (PBE and BLYP) are examined, while CCSDT \cite{shavitt2009many} results are used as reference. The basis set used is aug-cc-pVQZ. All energies are in kcal/mol.}
 \label{fig:Energy_C}
\end{figure}

In addition, the energies of C atom under varying external charged environments are tested. Fig \ref{fig:Energy_C} shows that the triplet-singlet energy gaps by KS seriously deviate  from the results of CCSDT, especially when the point charges are far away from the C atom. By comparison, HCKS significantly improves the performance of KS, in use of the same XC functionals. The occupations of the three p orbitals change gradually from $\{1,0,0\}$ to $\{1/3, 1/3, 1/3\}$ as the point charges move away from the C atom. Therefore, HCKS is able to capture the physical essence of multi-reference nature through dynamically varying occupations.

\begin{figure}[htbp]
 \centering
 \includegraphics[width=1\linewidth]{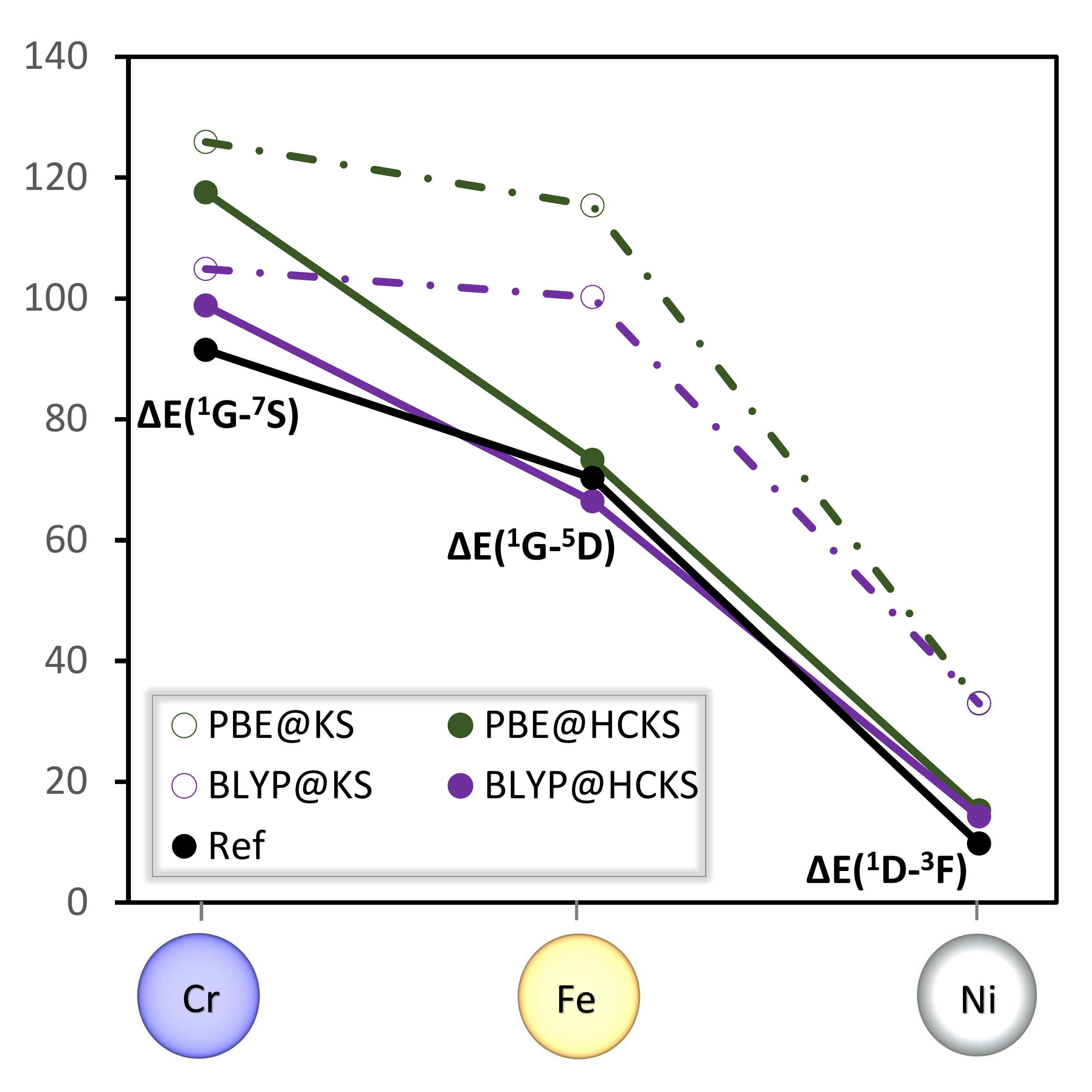}
 \caption{Energy gaps between high and low spin-states of transition metals. Septet-singlet, quintet-singlet, triplet-singlet energy gaps for Cr, Fe and Ni atoms respectively are calculated. KS and HCKS (or equivalently, RDMFT) in use of the same XC functionals (PBE and BLYP) are examined, while the experimental data \cite{Moore1952} are used as reference. The basis set used is def2-TZVP \cite{Weigend2005pccp}. All energies are in kcal/mol.}
 \label{fig:Energy_TM}
\end{figure}

Here test also some atoms of 3d transition metals, which are generally considered to be a big challenge to KS. Due to the partially filled 3d sub-shell and the near degeneracy of 4s and 3d sub-shells, these atoms or systems containing them often
have a plethora of low-lying degenerate and near-degenerate states, which make them much more difficult to correctly describe than main-group compounds. Fig \ref{fig:Energy_TM} shows that KS with both PBE and BLYP XC functionals overestimates the energy gaps between high and low spin-states of Cr, Fe and Ni, due to the lack of strong correlation in the low spin-states. In contrast, HCKS improves the performance in use of the same XC functionals. The fractional occupations for (nearly) degenerate orbitals render HCKS great potential for handling strong correlation. For example, PBE@HCKS converges to a set of spin-restricted orbitals for the singlet state of Ni,  with occupations for both $\alpha$ and $\beta$ spins being 0.916 for each 3d orbital and 0.420 for 4s orbital. This further proves that HCKS can improve the description of strong correlation while maintaining both spin and space symmetry.

$Conclusions.$---This letter presented a theory to unify the two theoretical frameworks of KS-DFT and RDMFT. This is achieved by HCKS that generalizes the KS orbitals to hypercomplex number systems. HCKS extends the search space for the density in KS-DFT to a space that is equivalent to natural spin orbitals with fractional occupations in RDMFT. The test on the singlet biradical C atom shows that HCKS can maintain both spin and space symmetry for the density with equally fractionally occupied p orbitals, which cannot be achieved by RKS. Besides, the calculation of energy gaps between high and low spin states, for both C atom under varying external charged environments and atoms of transition metals, demonstrates that HCKS is able to capture the multi-reference nature of strong correlation by dynamically varying fractional occupations, while KS in use of the same XC functionals cannot. Therefore, HCKS shows great potential to overcome the fundamental limitations of KS, which thus provides an alternative to the realization of DFT, and opens up new channels for the development and evaluation of approximate functionals. 

%
% acknowledgment
%
The Supplemental Material for this work is available on line \cite{footnote}. 
Support from the National Natural Science Foundation of China (Grant 22073049), the Natural Science Foundation of Tianjin City (20JCQNJC01760), and Fundamental Research Funds for the Central Universities (Nankai University: No. 63206008) is appreciated. Dedicated to the 100th anniversary of Chemistry at Nankai University.

\bibliographystyle{aip}
\bibliography{ref}

%\pagebreak
\onecolumngrid

\newcommand{\ti}{\Tilde}
\newcommand{\nl}{\nonumber \\}
\newcommand{\Sch}{Schr\"{o}dinger\ }
\newcommand{\Sec}[1]{Sec.\;\ref{#1}}
\newcommand{\App}[1]{Appendix\;\ref{#1}}
\newcommand{\be}{\begin{equation}}
\newcommand{\ee}{\end{equation}}
\newcommand{\bea}{\begin{eqnarray}}
\newcommand{\eea}{\end{eqnarray}}
\newcommand{\bsube}{\begin{subequations}}
\newcommand{\esube}{\end{subequations}}
\newcommand{\Eq}[1]{Eq.\,(\ref{#1})}
\newcommand{\Eqs}[1]{Eqs.\,(\ref{#1})}
\newcommand{\Fig}[1]{Fig.\,\ref{#1}}
\newcommand{\Tab}[1]{Table\,\ref{#1}}
\newcommand{\sub}[1]{$_{#1}$}
\newcommand{\upp}[1]{$^{#1}$}
\newcommand{\dg}{\dagger}
\newcommand{\la}{\langle}
\newcommand{\ra}{\rangle}
\newcommand{\kb}{\rangle\langle}
\newcommand{\w}{\omega}
\newcommand{\ep}{\epsilon}
\newcommand{\B}{\mbox{\tiny B}}
\newcommand{\bfGam}{\mbox{\boldmath $\Gamma$}}

%%
%\begin{document}
\vspace{10cm}

\draft

Supporting Information of
%
%\vspace{0.2cm}
\vspace{0.1cm}

\begin{center}
{\large \textbf{Unity of Kohn-Sham Density Functional Theory and Reduced Density Matrix Functional Theory}}
\vspace{0.2cm}

Neil Qiang Su

{\small
\emph{Department of Chemistry, Key Laboratory of Advanced Energy Materials Chemistry (Ministry of Education) and Renewable Energy Conversion and Storage Center (RECAST), Nankai University, Tianjin 300071, China\\}
\emph{\rm Email: nqsu@nankai.edu.cn}
}

\vspace{0.1cm}
%\today
%
\end{center}

\linespread{1.0}

%\vspace{0.2cm}

\tableofcontents

%%%%%%%%%%%%%%%%%%%
\renewcommand{\theequation}{S\arabic{equation}}
\renewcommand{\thetable}{S\arabic{table}}
\renewcommand{\thefigure}{S\arabic{figure}}
\setcounter{equation}{0}   
%%%%%%%%%%%%%%%%%%%%%

\section{Reduced Density Matrix Functional Theory}

In RDMFT, the electronic ground-state energy can be written as a functional of the one-body reduced density matrix (1-RDM) $\gamma_\sigma$ \cite{Gilbert1975prb,Levy1979pnas}
\begin{equation}
\label{seq:dmft}
E^{\rm{DMFT}}[\gamma_\sigma]=T[\gamma_\sigma]+E_{\rm{ext}}[\gamma_\sigma]+E_{\rm{H}}[\gamma_\sigma]+E_{\rm{XC}}[\gamma_\sigma],
\end{equation}
where the kinetic energy is 
\begin{equation}
\label{seq:T}
T[\gamma_\sigma]=-\frac{1}{2}\sum_{\sigma}^{\alpha,\beta}\int\nabla_{\br}^2\gamma_\sigma(\br,\br')|_{\br'=\br}d\br,
\end{equation}
the external energy is
\begin{equation}
\label{seq:Eext}
E_{\rm{ext}}[\gamma_\sigma]=\sum_{\sigma}^{\alpha,\beta}\int v_{\rm{ext}}(\br)\gamma_\sigma(\br,\br')|_{\br'=\br}d\br,
\end{equation}
the Coulomb energy is
\begin{equation}
\label{seq:EH}
E_{\rm{H}}[\gamma_\sigma]=\frac{1}{2}\sum_{\sigma\sigma'}^{\alpha,\beta}\int\frac{\gamma_\sigma(\br_1,\br'_1)\gamma_{\sigma'}(\br_2,\br'_2)|{^{\br'_1=\br_1}_{\br'_2=\br_2}}}{r_{12}}d\br_1d\br_2,
\end{equation}
and the exchange-correlation (XC) energy $E_{\rm{XC}}[\gamma_\sigma]$ is unknown. The total 1-RDM is
\begin{equation}
\label{seq:TRDM}
\gamma(\br,\br')=\gamma_\alpha(\br,\br')+\gamma_\beta(\br,\br')=N\sum_{\sigma_1,\cdots,\sigma_N}^{\alpha,\beta}\int d\br_2\cdots d\br_N\Psi^*(\br'\sigma_1,\bx_2,\cdots,\bx_N)\Psi(\br\sigma_1,\bx_2,\cdots,\bx_N),
\end{equation}
where $\bx_p=\br_p\sigma_p$ are combined spatial and spin coordinates. Diagonalization of 1-RDM $\gamma_\sigma$ generates the natural spin orbitals, $\{\psi_p^\sigma\}$, and their occupation numbers, $\{n_p^\sigma\}$, thus
\begin{equation}
\label{seq:aRDM}
\gamma_\sigma=\sum_{p=1}^K|\psi_p^\sigma\rangle n_p^\sigma \langle \psi_p^\sigma |; \gamma_\sigma(\br,\br')=\langle \br | \gamma_\sigma | \br' \rangle,
\end{equation}
where its diagonal element is the density,
\begin{equation}
\label{seq:rho_RDM}
 \rho_\sigma(\br)=\langle \br | \gamma_\sigma | \br \rangle=\sum_{p=1}^K n_p^\sigma |\psi_p^\sigma(\br)|^2.
\end{equation}
The ground-state energy can be obtained by minimizing Eq. \ref{seq:dmft} with respect to 1-RDM, or equivalently, by minimizing Eq. \ref{seq:dmft} with respect to both $\{\psi_p^\sigma\}$ and $\{n_p^\sigma\}$, subject to
\begin{equation}
\label{seq:cond}
\langle \psi_p^\sigma | \psi_q^\sigma \rangle=\delta_{pq}, 0\leq n_p^\sigma \leq 1, \sum_{p=1}^K n_p^\sigma=N_\sigma .
\end{equation}

Here, $\sigma$ is the spin index, it can be $\alpha$ or $\beta$; $N_\sigma$ is the electron number of $\sigma$ spin; $K$ is the dimension of basic functions, which should be not smaller than $N_\sigma$.

\section{Kohn-Sham Density Functional Theory}

In KS-DFT \cite{HK1964,KS1965}, the total energy is 
\begin{equation}
\label{seq:KS}
E^{\rm{KS}}[\rho_\sigma]=T_{\rm{s}}[\rho_\sigma]+E_{\rm{ext}}[\rho_\sigma]+E_{\rm{H}}[\rho_\sigma]+E_{\rm{XC}}[\rho_\sigma],
\end{equation}
where the kinetic energy is
\begin{equation}
\label{seq:Tks}
T_{\rm{s}}[\rho_\sigma]=-\frac{1}{2}\sum_{\sigma}^{\alpha,\beta}\int\nabla_{\br}^2\gamma_\sigma^s(\br,\br')|_{\br'=\br}d\br=-\frac{1}{2}\sum_{\sigma}^{\alpha,\beta}\sum_{k=1}^{N_\sigma}\langle \varphi_{k}^{\sigma} | \nabla^2 | \varphi_{k}^{\sigma} \rangle,
\end{equation}
the external energy is
\begin{equation}
\label{seq:Eextks}
E_{\rm{ext}}[\rho_\sigma]=\sum_{\sigma}^{\alpha,\beta}\int v_{\rm{ext}}(\br)\rho_\sigma(\br)d\br,
\end{equation}
the Coulomb energy is
\begin{equation}
\label{seq:EHks}
E_{\rm{H}}[\rho_\sigma]=\frac{1}{2}\sum_{\sigma\sigma'}^{\alpha,\beta}\int\frac{\rho_\sigma(\br_1)\rho_{\sigma'}(\br_2)}{r_{12}}d\br_1d\br_2,
\end{equation}
while the form of $E_{\rm{XC}}[\rho_\sigma]$ is unknown. 

$\gamma_\sigma^s$ is 1-RDM of the noninteracting reference system, which can be established by occupied orbitals,
\begin{equation}
\label{seq:aRDMks}
\gamma_\sigma^s=\sum_{k=1}^{N_\sigma}|\psi_k^\sigma\rangle \langle \psi_k^\sigma |; \gamma_\sigma^s(\br,\br')=\langle \br | \gamma_\sigma^s | \br' \rangle,
\end{equation}
and the density is
\begin{equation}
\label{seq:rhoks}
\rho_\sigma(\br)=\langle \br | \gamma_\sigma^s | \br \rangle=\sum_{k=1}^{N_\sigma} |\varphi_k^\sigma(\br)|^2.
\end{equation}
Here $\{\varphi_p^\sigma\}$ are the KS orbitals, which are a set of orthonormal orbitals that span the space of dimension $K$, 
\begin{equation}
\label{seq:orthoks}
\langle \varphi_p^\sigma | \varphi_q^\sigma \rangle=\delta_{pq},
\end{equation}
thereby
\begin{equation}
\label{seq:iden_ks}
1_K=\sum_{p=1}^K |\varphi_p^\sigma\rangle\langle\varphi_p^\sigma|.
\end{equation}
Here $1_K$ represents the identity operator of the space.

\section{Hypercomplex numbers}

Hypercomplex \cite{HC1989} and the theory of Clifford algebra \cite{CA2019} that generalizes real numbers, complex numbers to quanternions, octonions, and other hypercomplex numbers are introduced here.

Clifford algebra is a unital associative algebra generated by a vector space with a quadratic form, which has important applications in a variety of fields including theoretical physics. In a Clifford algebra, given a basis of dimension $n$, $\{e_1, e_2, \cdots, e_n\}$, such that \cite{CA2019}
\begin{equation}
\label{seq:ca}
e_\mu^2=-1; e_\mu e_\nu=-e_\nu e_\mu,
\end{equation}
further imposing closure under multiplication generates a multivector space spanned by a basis of the $2^n$ distinct products of $\{e_1, e_2, \cdots, e_n\}$, i.e. $\{1, e_1, e_2, \cdots, e_1 e_2, \cdots, e_1 e_2 e_3, \cdots\}$. These distinct products form the basis of a hypercomplex number system. Unlike $\{e_1, e_2, \cdots, e_n\}$, the remaining elements in this basis need not anti-commute, depending on how many simple exchanges must be carried out for the swap. 

Here we consider the following hypercomplex number
\begin{equation}
\label{seq:hc}
C=\sum_{\mu=0}^{n}c^\mu e_\mu=c^0\cdot1+\sum_{\mu=1}^{n}c^\mu e_\mu,
\end{equation}
where $e_0=1$, and the coefficients $\{c^\mu\}$ are real. The corresponding conjugate hypercomplex number of Eq. \ref{seq:hc} is
\begin{equation}
\label{seq:chc}
\bar{C}=c^0\cdot1-\sum_{\mu=1}^{n}c^\mu e_\mu.
\end{equation}

\section{Hypercomplex Kohn-Sham Theory}

Now, we generalize the KS orbitals to hypercomplex number systems,
\begin{equation}
\label{seq:varphi_hc}
\varphi_p^\sigma(\br)=\sum_{\mu=0}^{n}\phi_{p}^{\sigma,\mu}(\br) e_\mu.
\end{equation}
The density of Eq. \ref{seq:rhoks} becomes
\begin{equation}
\label{seq:rho_hc}
\rho_\sigma(\br)=\sum_{k=1}^{N_\sigma}\sum_{\mu=0}^{n}[\phi_{k}^{\sigma,\mu}(\br)]^2,
\end{equation}
and the kinetic energy of Eq. \ref{seq:Tks} becomes
\begin{equation}
\label{seq:T_hc}
T_{\rm{s}}[\rho_\sigma]=-\frac{1}{2}\sum_{\sigma}^{\alpha,\beta}\sum_{k=1}^{N_\sigma}\sum_{\mu=0}^{n} \langle\phi_{k}^{\sigma,\mu} |\nabla^2 | \phi_{k}^{\sigma,\mu} \rangle.
\end{equation}

Without loss of generality, we expand $\{\phi_{p}^{\sigma,\mu}\}$ on a set of orthonormal functions $\{\chi_p\}$, i.e.
\begin{equation}
\label{seq:chi2phi}
\phi_{p}^{\sigma,\mu}(\br)=\sum_{q=1}^K \chi_q(\br)V_{pq}^{\sigma,\mu}.
\end{equation}
Here $\{{\rm{V}}^{\sigma,\mu}\}$ are a set of $K\times K$ matrices. Eq. \ref{seq:orthoks} becomes
\begin{equation}
\label{seq:cond1_hc}
\sum_{\mu=0}^{n} {\rm{V}}^{\sigma,\mu}{\rm{V}}^{\sigma,\mu T}={\rm{I}}_K,
\end{equation}
and
\begin{equation}
\label{seq:cond2_hc}
{\rm{V}}^{\sigma,\mu}{\rm{V}}^{\sigma,\nu T}={\rm{V}}^{\sigma,\nu}{\rm{V}}^{\sigma,\mu T}.
\end{equation}
Eq. \ref{seq:iden_ks} leads to $\langle \chi_p | \chi_q \rangle =\sum_{r=1}^K \langle \chi_p |\varphi_r^\sigma\rangle\langle\varphi_r^\sigma|\chi_q \rangle=\delta_{pq}$, such that 
\begin{equation}
\label{seq:cond3_hc}
\sum_{\mu=0}^{n} {\rm{V}}^{\sigma,\mu T}{\rm{V}}^{\sigma,\mu}={\rm{I}}_K,
\end{equation}
and
\begin{equation}
\label{seq:cond4_hc}
{\rm{V}}^{\sigma,\mu T}{\rm{V}}^{\sigma,\nu}={\rm{V}}^{\sigma,\nu T}{\rm{V}}^{\sigma,\mu}.
\end{equation}
${\rm{I}}_K$ denotes the $K\times K$ identity matrix.

Here we define two $K\times K$ symmetric matrices ${\rm{D}}^\sigma$ and ${\rm{E}}^\sigma$, 
\begin{equation}
\label{seq:XandY}
{\rm{D}}^\sigma=\sum_{\mu=0}^{n} {\rm{V}}^{\sigma,\mu T}{\rm{I}}_K^{N_\sigma}{\rm{V}}^{\sigma,\mu}; {\rm{E}}^\sigma=\sum_{\mu=0}^{n} {\rm{V}}^{\sigma,\mu T}{\rm{I}}_K{\rm{V}}^{\sigma,\mu T}=\sum_{\mu=0}^{n} {\rm{V}}^{\sigma,\mu T}{\rm{V}}^{\sigma,\mu T}={\rm{I}}_K.
\end{equation}
${\rm{I}}_K^{N_\sigma}$ is a $K \times K$ diagonal matrix, with the first $N_\sigma$ diagonal elements being 1 and the rest being 0. 
${\rm{D}}^\sigma$ can be diagonalized by an orthogonal matrix ${\rm{U}}^\sigma$, thus
\begin{equation}
\label{seq:ULU}
{\rm{D}}^\sigma={\rm{U}}^\sigma {\rm{\Lambda}}^\sigma {\rm{U}}^{\sigma T},
\end{equation}
where ${\rm{\Lambda}}^\sigma$ is a diagonal matrix with diagonal elements $\{\lambda_p^\sigma\}$ being the eigenvalues of ${\rm{D}}^\sigma$. The positive semi-definiteness of ${\rm{D}}^\sigma$ guarantees that all the eigenvalues are not negative, with the maximal eigenvalue smaller than that of ${\rm{E}}^\sigma$. Thereby,
\begin{equation}
\label{seq:occ_cond}
0 \leq \lambda_p^\sigma \leq 1; \sum_{p=1}^K \lambda_p^\sigma={\rm{Tr}}\left({\rm{D}}^\sigma\right)={\rm{Tr}}\left(\sum_{\mu=0}^{n} {\rm{V}}^{\sigma,\mu T}{\rm{I}}_K^{N_\sigma}{\rm{V}}^{\sigma,\mu}\right)={\rm{Tr}}\left({\rm{I}}_K^{N_\sigma}\sum_{\mu=0}^{n} {\rm{V}}^{\sigma,\mu}{\rm{V}}^{\sigma,\mu T}\right)=N_\sigma.
\end{equation}
The density and kinetic energy can be formulated as
\begin{equation}
\label{seq:Vrho}
\rho_\sigma(\br)=\sum_{p,q=1}^K  \chi_p(\br)D_{pq}^{\sigma} \chi_q(\br)=\sum_{p=1}^{K} \lambda_{p}^{\sigma} |\chi_p^\sigma(\br)|^2,
\end{equation}
and
\begin{equation}
\label{seq:VT}
T[\rho_\sigma]=-\frac{1}{2}\sum_{\sigma}^{\alpha,\beta}\sum_{p,q=1}^K D_{pq}^{\sigma} \langle\chi_p|\nabla^2|\chi_q\rangle=-\frac{1}{2}\sum_{\sigma}^{\alpha,\beta}\sum_{p=1}^{K} \lambda_{p}^{\sigma} \langle \chi_p^\sigma | \nabla^2 | \chi_p^{\sigma} \rangle,
\end{equation}
where $\chi_p^\sigma(\br)=\sum_{q=1}^K \chi_q(\br)U_{qp}^\sigma$.

For $K\leq n+1$, to show that $\{\lambda_p^\sigma\}$ can be any values that satisfy Eq. \ref{seq:occ_cond}, here we construct a special set of orbitals $\{\varphi_p^\sigma\}$ on any given set of orthonormal functions $\{\chi_p^\sigma\}$, which take the following form
\begin{equation}
\label{seq:varphi2}
\varphi_p^\sigma(\br)=\sum_{\mu=0}^{n}\phi^{\sigma,\mu}_p e_\mu=\sum_{\mu=0}^{K-1} \chi_{\mu+1}^\sigma V_{p\mu+1}^{\sigma,\mu} e_{\mu},
\end{equation}
i.e. the components corresponding to the same $e_\mu$ in all the orbitals $\{\varphi_p^\sigma\}$ are formed by only the function $\chi_{\mu+1}^\sigma$, thereby
\begin{align}
\label{seq:V2}
V_{pq}^{\sigma,\mu}=\begin{cases}
W_{pq}^\sigma, & q=\mu+1\\
0, & q\neq \mu+1,
\end{cases}
\end{align}
where ${\rm{W}}^\sigma$ is a $K \times K$ matrix. Now the conditions of Eqs. \ref{seq:cond1_hc}-\ref{seq:cond4_hc} become
\begin{equation}
\label{seq:cond11_hc}
{\rm{W}}^{\sigma}{\rm{W}}^{\sigma T}={\rm{I}}_K,
\end{equation}
\begin{equation}
\label{seq:cond21_hc}
W_{k,\mu+1}W_{l,\nu+1}\delta_{\mu\nu}=W_{k,\nu+1}W_{l,\mu+1}\delta_{\mu\nu},
\end{equation}
\begin{equation}
\label{seq:cond31_hc}
[{\rm{W}}^{\sigma T}{\rm{W}}^{\sigma}]_{\mu+1,\mu+1}=1,
\end{equation}
\begin{equation}
\label{seq:cond41_hc}
[{\rm{W}}^{\sigma T}{\rm{W}}^{\sigma}]_{\mu+1,\nu+1}=[{\rm{W}}^{\sigma T}{\rm{W}}^{\sigma}]_{\nu+1,\mu+1}.
\end{equation}
Eqs. \ref{seq:cond11_hc}-\ref{seq:cond41_hc} can be satisfied once ${\rm{W}}^\sigma$ is an orthogonal matrix. Now, the density and the kinetic energy becomes
\begin{equation}
\label{seq:Wrho}
\rho_\sigma(\br)=\sum_{p=1}^K \sum_{k=1}^{N_\sigma}(W_{kp}^{\sigma})^2  |\chi_p^\sigma(\br)|^2,
\end{equation}
and
\begin{equation}
\label{seq:WT}
T[\rho_\sigma]=-\frac{1}{2}\sum_{\sigma}^{\alpha,\beta}\sum_{p=1}^K \sum_{k=1}^{N_\sigma}(W_{kp}^{\sigma})^2 \langle\chi_p^\sigma | \nabla^2 | \chi_p^\sigma \rangle.
\end{equation}
Hence $\lambda_p^\sigma = \sum_{k=1}^{N_\sigma}(W_{kp}^{\sigma})^2$. Because ${\rm{W}}^\sigma$ can be any unitary matrix, $\{\lambda_p^\sigma\}$ can be any values subject to 
\begin{equation}
0\leq \lambda_p^\sigma \leq 1; \sum_{p=1}^K \lambda_p^\sigma =N_\sigma.
\end{equation}

When $n=0$, the orbitals of Eq. \ref{seq:varphi_hc} are real and the hypercomplex KS (HCKS) reduces the conventional KS method; when $n=1$, the orbitals of Eq. \ref{seq:varphi_hc} are complex, which are used in the complex, spin-restricted KS (CRKS) method \cite{Lee2019prl}. Take the singlet state of C atom as an example to compare KS, CRKS and HCKS. The three methods all lead to fully occupied 1s and 2s sub-shells that are lower in energy than the 2p sub-shell. For the 2p sub-shell, the spin-restricted KS (RKS) has one doubly occupied p orbital; CRKS leads to two half filled p orbitals as Eq. \ref{seq:varphi_hc} with $n=1$ has only two components for each occupied orbitals; HCKS automatically converges to the spin-restricted solution, with each p orbital 1/3 filled. Therefore, the density by HCKS can maintain both spin and space symmetry. Note that in most cases especially when the energy gap of the frontier orbitals is big, HCKS in use of commonly used functionals gives the same results as KS.

%\bibliographystyle{aip}
%\bibliography{ref2}

\end{document}